\documentclass[12pt]{article}

\usepackage[english]{babel}
\usepackage[usenames]{color}
\usepackage[cp1250]{inputenc}
\usepackage{amsfonts}
\usepackage{amsthm}
\usepackage{graphicx}
\usepackage{epsfig}

\theoremstyle{plain}

\begin{document}
\newcommand{\bea}{\begin{eqnarray}}
\newcommand{\eea}{\end{eqnarray}}
\newcommand{\be}{\begin{equation}}
\newcommand{\ee}{\end{equation}}
\newcommand{\beas}{\begin{eqnarray*}}
\newcommand{\eeas}{\end{eqnarray*}}
\newcommand{\bs}{\backslash}
\newcommand{\bc}{\begin{center}}
\newcommand{\ec}{\end{center}}

\title{Embedding grayscale halftone pictures in QR Codes \\ using Correction Trees}
\author{Jarek Duda}

\date{\it \footnotesize Jagiellonian University, Cracow, Poland, \\
\textit{email:} dudaj@interia.pl}
\maketitle

\begin{abstract}
Barcodes like QR Codes have made that encoded messages have entered our everyday life, what suggests to attach them a second layer of information: directly available to human receiver for informational or marketing purposes. We will discuss a general problem of using codes with chosen statistical constrains, for example reproducing given grayscale picture using halftone technique. If both sender and receiver know these constrains, the optimal capacity can be easily approached by entropy coder. The problem is that this time only the sender knows them - we will refer to these scenarios as constrained coding. Kuznetsov and Tsybakov problem in which only the sender knows which bits are fixed can be seen as a special case, surprisingly approaching the same capacity as if both sides would know the constrains. We will analyze Correction Trees to approach analogous capacity in the general case - use weaker: statistical constrains, what allows to apply them to all bits. Finding satisfying coding is similar to finding the proper correction in error correction problem, but instead of single ensured possibility, there are now statistically expected some. While in standard steganography we hide information in the least important bits, this time we create codes resembling given picture - hide information in the freedom of realizing grayness by black and white pixels using halftone technique. We will also discuss combining with error correction and application to rate distortion problem.
\end{abstract}


\section{Introduction}
The lack of knowledge of damage positions makes that the Binary Symmetric Channel (BSC, each bit has $p_b$ probability of being flipped) has relatively low rates: limited by very difficult to approach $1-h(p_b)$ rate of Shannon's Noisy Channel Coding Theorem ($h(p):=-p \lg(p)-(1-p)\lg(1-p),\ \lg\equiv \log_2$). This limit can be intuitively seen that if we would additionally know the bit flip positions, what is worth $h(p_b)$ per bit, the rate would be 1. In contrast, if both sides would know which $p_e$ of bits are somehow erased/lost, they could easily use the remaining bits to achieve the maximal: $1-p_e$ rate. In Erasure Channel only the receiver knows the positions of erased bits, but still we can relatively easily get close to the $1-p_e$ rate of using only the remaining bits.

\begin{table}[b!]
    \centering
        \includegraphics{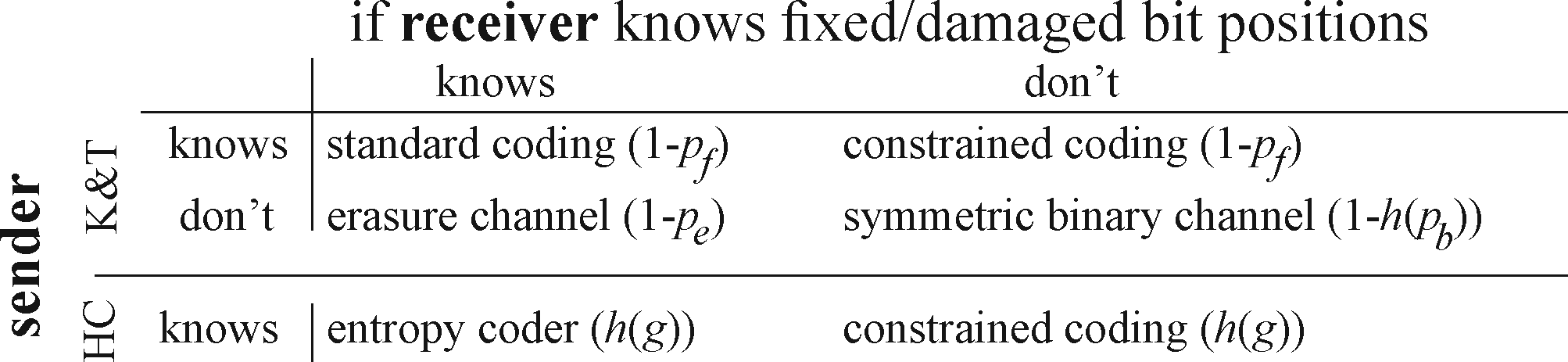}
        \caption{Basic scenarios and their rate limits for Kuznetsov and Tsybakov-like case (K\&T) - some bit are fixed/damaged and homogeneous contrast case (HC) - all bits are intended to have $(g,1-g)$ probability distribution (or opposite). $p_f$ is probability of bit being fixed, $p_e$ of being erased(lost), $p_b$ of bit flips in unknown positions.}
       \label{tab}
\end{table}

Kuznetsov and Tsybakov \cite{init} have asked kind of opposite question: what if the only side knowing the locations of damages is the sender? This time by damages we mean arbitrarily fixing some bits, like fixing some picture in a region of QR code. The first answer is that we could use error correction method for this purpose - encode the message with redundancy, then eventually modify (damage) the fixed bits - we will refer to this approach as \emph{damaged ECC}. The receiver does not know the positions of the fixed bits, so it is BSC scenario from his perspective. Half of these bits are expected to accidently have already the required values, so this way we could get maximally $1-h(p_f/2)$ rate if random $p_f$ of bits are fixed. For example for $p_f=0.5$, $1-h(p_f/2)\approx 0.1887$, while $1-p_f=0.5$ is 2.65 times larger.

Surprisingly, the limit turns out to be as for the Erasure Channel - we can use nearly all undamaged bits, for example getting nearly 2.65 larger rate in the $p_f=0.5$ case (because of difficulty to approach BSC limit, in practice even more). To understand that this improvement is possible, observe that in damaged ECC approach the receiver additionally obtains (half of) positions of fixed bits while performing the correction - intuitively we unintentionally attach this unnecessary information.

So how to remove this "damage locations information"? We can use the freedom of choosing the exact encoding message, such that it just "accidently" agrees with the constrains - thanks of it, the receiver is not able to distinguish the intentional bits from the fixed ones. We will do it by using regularly distributed \emph{freedom bit(s)}: which will be just discarded while decoding, but gives a freedom while choosing the exact encoding sequence. For example we can try to use only zeros for these freedom bits, but when produced block does not satisfy the constrains, we try out different values for these bits. So intuitively we search for a satisfying path in a tree of potentially $2^{k}$ leaves: different ways to encode the message, where $k$ is the total number of freedom bits up to the current position - if this number is essentially larger than the number of fixed bits there, there is large probability that an encoding fulfilling these constrains is included. After that, decoding is straightforward - most of the computational cost is paid by the encoder, what is very convenient e.g. for 2D codes application.\\

The original motivation of the problem was Write-Once-Memory which fixes to the value after being used. Relatively recent 2D codes like QR Codes \cite{qr} rapidly entering everyday life, bring another direct application - while standard codes are usually designed to be processed only by specialized algorithms and so should use the most efficient $P(0)=P(1)=1/2$ bit distribution, the fact that our brains also try to directly process 2D codes suggests to construct them to simultaneously deliver also some information in visual: human-friendly layer.

And so we can currently meet 2D codes with embedded some simple picture, usually by just flipping some bits of the original code (damaged ECC approach). As the current standards are not optimized for this purpose, the design process uses included redundancy - greatly reducing correction capabilities of such already damaged code and most importantly: using this informational channel in very ineffective way (by unintentionally attaching damage locations as discussed above). So one direction of evolution of 2D codes might be to optimize them to also embed additional e.g. visual information. We will discuss and analyze how to use Correction Trees (\cite{me},\cite{cortre}, simulator: \cite{cortre1}) for this purpose, which can be seen as extended concept of sequential decoding for Convolutional Codes - some major improvements has made it alternative to the state of art error correction methods. It is also very convenient to allow for freedom to efficiently search for encoding message satisfying required constrains. Finally, while in the original method the decoder searches the tree of possibilities to find the proper correction of the received message, now additionally the encoder searches the tree of possibilities to find encoding message fulfilling the constrains - connecting both features in nearly optimal way.\\

The application for 2D codes also suggests a generalization of the original problem: allow not only to fix some bits, but also to choose statistics for them, for example to simulate grayscale like in the halftone technique - using large enough resolution we could get 2D codes which look like a chosen grayscale picture, in which bits (black and white pixels) emerge while getting closer. While fixed bits can no longer contain encoded information (only visual), the freedom of grayness realization allows to relatively cheaply apply statistical constrains to all of pixels.

Communication through messages of chosen bit statistics can be easily achieved by using an entropy coder like Arithmetic Coding or Asymmetric Numeral Systems(\cite{me}): treating 0/1 as symbols of chosen probability. However, it requires that both sides know these statistical constrains, which can vary locally - corresponds to situation that both sides know locations of damaged bits. We will see that if only the sender knows the desired local statistics, the channel capacity can remain nearly the same. It might be useful if for some reasons the channel prefers e.g. some varying bit statistics, which is known only to the sender.

Example of such purposes can be cryptographic - the use of the original defective memory problem was already considered for steganographic applications (\cite{steg}). Presented generalization additionally allows for example to transmit information in faking a noise of some process/device - in situations that it varies in time and only the sender can effectively estimate its statistics.

\section{Homogeneous contrast constrained coding limits}
Thanks to some pixel ordering (discussed later), we can represent a bit sequence as black ('1') and white ('0') rectangular picture of relatively low resolution: 2D code. Let us define \emph{grayness} $g\in [0,1]$ of a pixel as probability of using black color ('1') for this position in our coding - pixels fixed to white or black have grayness 0 or 1 correspondingly. Standard optimal coding techniques should usually produce typical bit sequences: with equal probabilities of digits and without correlations - corresponding to $1/2$ grayness.

\begin{figure}[b!]
    \centering
        \includegraphics{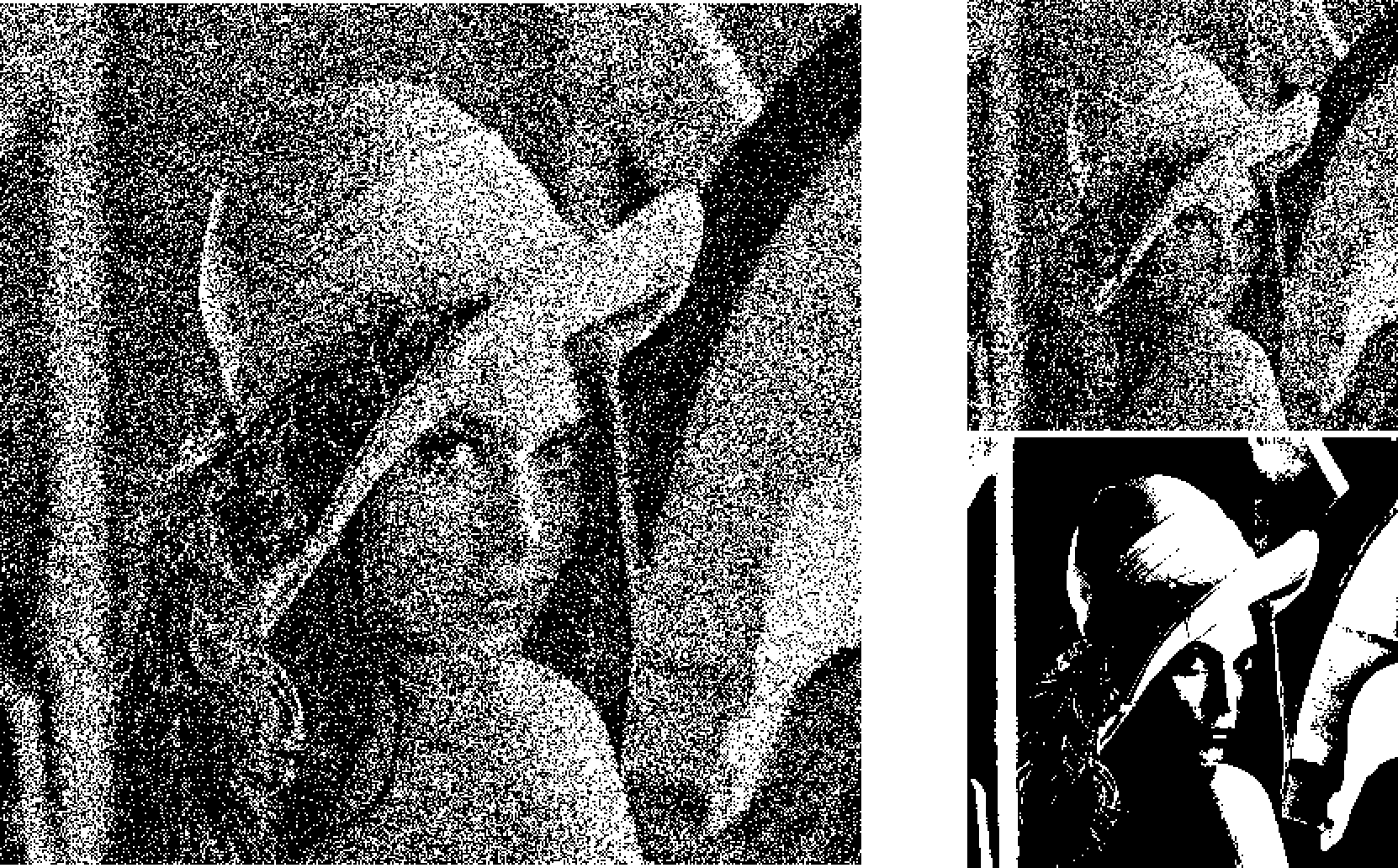}
        \caption{\emph{Lena}-like looking codes of 512x512 (left) and 256x256 (right top) resolutions. Rate limit for this picture is about 0.822 (average $h(g)$ over all pixels) - the visual aspect may cost only 18\% rate reduction. For comparison, the bottom right picture contains the standard way of making 256x256 picture black and white: $g<1/2$ pixels become white and the rest become black. While it has much better contrast, it carries no additional information about grayness and does not allow to hide any additional message.}
       \label{lena}
\end{figure}
The final goal is to generate codes for halftonig a given grayscale picture, like in Fig. \ref{lena}: encode such that probability of '1' for each pixel is defined by grayness of corresponding pixel of the chosen picture. So for a given message and a grayscale picture (GP), we want to encode a message as a halftone picture (HP). If both encoder and decoder know the GP, we could just use entropy coder - treating the pixels as symbols of the chosen probability distribution (grayness). In such case, pixel of $g$ grayness carries asymptotically $h(g)$ bits of information, where
\be h(p):=-p \lg(p)-(1-p)\lg(1-p)\qquad\qquad\qquad\textrm{is Shannon's entropy, }\lg\equiv \log_2\ee
So accordingly to the source coding theorem, the total amount of information we should be able to store in such halftone picture is a bit smaller than the sum of $h(g)$ over all pixels. The problem is that decoder usually do not know the GP - surprisingly we can still store nearly the same amount of information in this case. This practical possibility is suggested by the Kuznetsov and Tsybakov problem which can be seen as a special case here: as choosing grayness of some pixels to 0 or 1 (fixing their values) and to 1/2 for the rest of pixels for maximal informational content.\\

\begin{figure}[b!]
    \centering
        \includegraphics{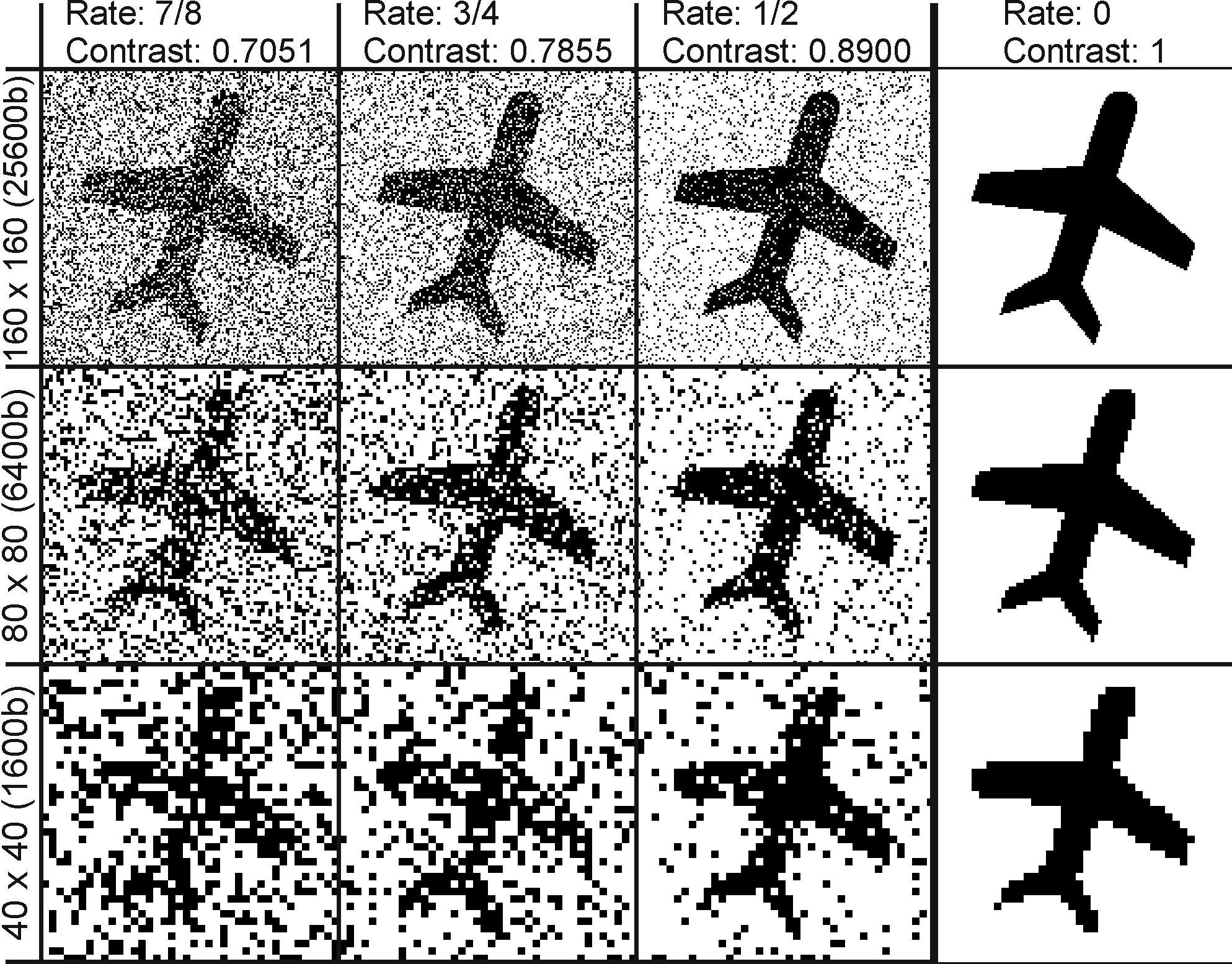}
        \caption{Examples of homogeneous contrast codes for three different resolutions and three different rates with maximal possible contrast. For example the central halftone picture stores $800\cdot 3/4=600$ bytes - encoding the visual structure costs 200 bytes (literally, as we will see in Section \ref{ratdis}). }
       \label{ex}
\end{figure}

This possibility can be obtained by using the huge freedom of choosing the exact code for our message. So imagine we attach some \emph{freedom bits} to the original message - the encoder can choose them freely, while the decoder will just discard them - we can assume that for each choice of these freedom bits, we get a different typical 0/1 sequence (Pr(0)=Pr(1)=1/2, no correlations) corresponding to the same message. Specifically, if $q$ (\emph{freedom level}) of $N$ bits of the message are freedom bits, we have $2^{Nq}$ possibilities to choose the exact encoding sequence. The trick is that this huge space of possibilities allows to choose sequences of extremely small probability - fulfilling assumed statistics (GP). In practice these freedom bits will be distributed uniformly between blocks of data (for example 1 freedom bit per 8 bit block corresponds to 7/8 rate coding). By developing the tree of possibilities we will find the most appropriate codings.

Let us start our considerations with kind of opposite side to the Kuznetsov and Tsybakov case: in which we want all bits to contain the same average amount of information. The freedom allows to shift the maximal information 1/2 grayness case symmetrically to both directions: dark areas of greyness $g>1/2$ and light areas of greyness $1-g<1/2$. So in this case we start with a black and white picture and encode information in added noise - it can be seen as a typical steganographic scenario, in which we for example want to encode information in a smallest possible disturbance of the least important bits of a picture using multiple bits per pixel. We will call it \emph{homogeneous contrast} case here of $g>1/2$ contrast - examples of possibilities of such codes can be seen in Fig. \ref{ex}.\\

The question is: what is the largest contrast available for given freedom level $q$? For $q=0$ (rate 1) all pixels have $g=1/2$, while for $q=1$ the rate is 0 so HP should be just the original black and white GP: the contrast is 1.

Imagine first we would like to obtain $g$ greyness level for all $N$ bits - make that about $gN$ of them are '1'. Probability of accidently obtaining it with typical sequences is
\be \frac{1}{2^N} {N \choose gN} \approx \frac{1}{2^N} 2^{Nh(g)}=\frac{1}{2^{N(1-h(g))}} \ee
so checking essentially more than $2^{N(1-h(g))}$ random sequences, with large probability we will accidently get this greyness level - using freedom level $q > 1-h(g)$ makes that asymptotically almost surely we will find such sequence. The remaining $(1-q)N$ bits are used to store the information (payload bits), so in this case the rate is $1-q<h(g)$, what is exactly the Shannon limit for storing information in symbols of $(g,1-g)$ probability distribution - we would get this limit simply using entropy coder, but it would require that decoder also knows the probability distribution ($g$). This time only the sender knows it and the receiver decodes the message as being standard $\mathrm{Pr}(0)=\mathrm{Pr}(1)=1/2$ bit sequence. The cost is the search through size $2^{N(1-h(g))}$ space of possibilities, what seems completely impractical. However, we will see that relatively cheap approximations allow to get very close to this limit, like considering some fixed number of possibilities up to some position and successfully shifting this ensemble.

To get from this constant grayness to homogeneous contrast case (with chosen light and dark pixels), instead of requiring that the number of black pixels is approximately $gN$, we would like that "the number of black pixels in light area + the number of white pixels in dark area" is the smallest possible, or equivalently: approximately $gN$. For more different greyness levels we can treat them separately, so finally we see that the asymptotic limit of information we can store in such halftone picture is the sum of $h(g)$ over all pixels, getting version of Shannon's source coding theorem in which the receiver does not need to know symbol probability distribution.\\

\begin{figure}[b!]
    \centering
        \includegraphics{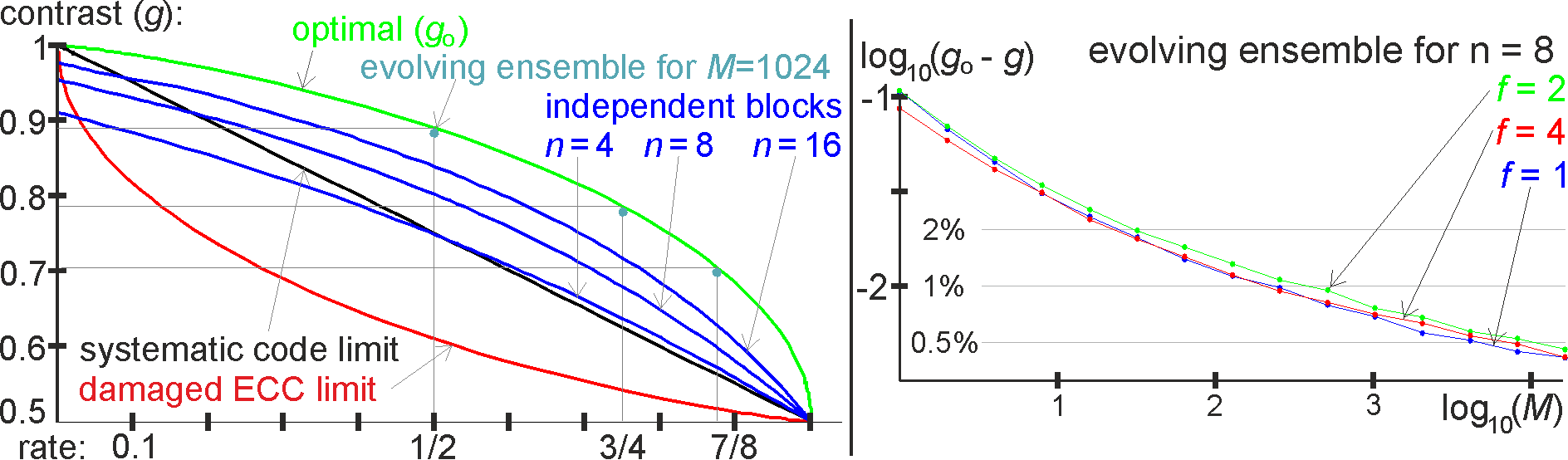}
        \caption{Left: limits for maximal homogeneous contrast for given rates and different approaches. Three points are from evolving ensemble for $n=8,\ M=1000$. Right: approaching the optimum by evolving ensemble approach.}
       \label{comp}
\end{figure}

Before going to practical approaches, let us compare this limit with standard ones. The current way to obtain chosen patterns embedded in 2D code is mainly using redundancy attached for error correction - by damaging the code in a convenient way. One problem of such approach (we will call \emph{damaged ECC}) is reduction of correction capabilities. Apart from this issue, let us ask how effective this approach can be for discussed purpose alone. To obtain $g$ grayness/contrast from the initial $1/2$, we would need to change (damage) on average $|1/2-g|$ of bits. Shannon's limit to handle this level of bit flips of positions unknown to receiver (BSC) is rate $1-h(|1/2-g|)$. From Fig. \ref{comp} we see it is much worse than the optimum, for example providing about 16 times smaller channel capacity for 0.9 contrast - such application is extreme waste of channel capacity.

Let us also compare it with boundary for \emph{systematic codes}: in which we directly store the message as some transmitted bits. Optimally used, these bits have grayness 1/2 and we can manipulate the rest of them to obtain the required statistics. Fixing $q$ of them to 0, we would get $(1-q)/2$ grayness, so to get $g$ greyness/contrast the rate would be $1-2|g-1/2|$. While it is much better than for damaged ECC case, it is still far from the optimum.
\section{Correction Trees for homogeneous contrast case}
We will now consider practical way to get close to the limit for homogeneous contrast case by splitting the process of choosing suitable freedom bits into single choices.

We will construct the code from fixed length ($n$) bit blocks in which we place some fixed number ($qn=f<n$) of freedom bits which can be freely chosen and will be discarded during decoding - the rest $k:=n-f$ bits will be called payload bits as containing the message we would like to store, so the rate is $k/n$. As example we will use $n=8,\ f=1$ (the rate is $7/8$): the code is built of bytes containing 7 bits of the original message, which can be chosen in $2^f=2$ ways. Building the code directly from such blocks would easily allow to reach the systematic code limit $(1-2|1/2-g|)$. To get nearer the optimum, we just have to use nonsystematic codes - somehow process such blocks to make payload bits no longer directly accessible.\\

Let us consider \emph{independent block} case first - that a block transformation is applied independently to each block, to see it is essential to intuitively connect their freedom. For unique decoding the transformation should be a bijection in the space of possible block values ($t:\{0,2^n-1\}\to \{0,2^n-1\}$) - it can be chosen randomly, for example using a pseudorandom number generator initialized with a cryptographic key. The question is what is the lowest grayness we can achieve this way, what naturally translates into the highest possible contrast for the homogeneous contrast case by reversing the condition for dark areas.

The freedom allows us to choose the best of $2^f$ length $n$ sequences - having the smallest number of '1's. So the question is the expected value of minimum of $2^f$ independent ${n\choose i}/2^n$ binomial distributions. The minimum is larger than a value if all variables are larger - its distribution can be simply expressed by cumulative distribution function (CDF):
$$CDF_{\textrm{minimum of $m$ variables $x$}}=1-(1-CDF_x)^m$$
Finally the expected minimal number of '1' for example for $n=4$ case is $(5^m+11^m+15^m+1)/16^m$ and we can get analogous formulas with CDF coefficients for different $n$. Substituting $m=2^f$, this formula provides the maximal contrast for given $q=f/n$ freedom level for independent block case. It allows to draw continuous plots, but it is difficult to interpret it for not natural $m$. Plots for $n=4,8,16$ are drawn in Fig. \ref{comp} - for $n\geq 4$ it can be better than for systematic coding, but is still far from the optimum. \\

Generally, as for error correction, the longer blocks the better - like Correction Trees connects redundancy of blocks, we will now connect their freedom to treat the whole message as single data sequence to be able to approach the optimum. For this purpose we can use a state connecting both freedom and redundancy. So let us call $x$ the preprocessed length $n$ bit block: $k$ payload bits, $f$ freedom bits. We would like to encode it as $y$ bit block. For independent blocks we have used $y=t(x)$ for some random bijection $t$. Including the internal state $s$, we would like a transformation function to produce $y$ and a new state while encoding: $(x,s)\to (y,s')$. While decoding, knowing $s$ and $y$ we should be able to determine $x$ in unique way. We can obtain it in simple and quick way by modifying the systematic coding from \cite{cortre} to nonsystematic one, by the way simultaneously allowing to include redundancy bits for error correction as we will discuss later: use tabled transition function $t:\{0,2^n-1\}\to \{0,2^N-1\}$, where $N$ is the size of the state, such that $t$ is bijection on the first $n$ bits. As discussed in \cite{cortre}, for effective use of all bits of the state, we should construct $t$ as concatenation of succeeding different pseudorandom bijections, like of 8 in $n=8$, $N=64$ standard case. Now the encoding step is: $y$ as XOR of the youngest bits of $t(x)$ and $s$, $s'$ as cyclically shifted $(s \textrm{ XOR }t(x))$.

The question is: what size of $s$ should we use? We could use short one like in Convolutional Codes. One of reasons for long state like $64$ bits in \cite{cortre} is for combining with error correction - providing much better performance. For the current purposes, short state should definitely not be used for strong constrains like in the Kuznetsov and Tsybakov case: the small space of states would have essential probability to not include a state fulfilling the constrains.\\

So let us imagine we produce a sequence of $y$ data blocks this way - the current situation is connected with the previous ones through the state. It makes the natural way to think of the space of possible codes as a tree: in which each node has potentially $2^f$ children.

We would like to elongate the looking promising paths - the question is how to do it? A natural way is to consider some number ($M$) of the most promising up to given position, then elongate all of them by a single bit block and choose only the most promising $M$ of these $2^f M$ to the next step - we will refer to this approach as \emph{evolving ensemble}. Intuitively, increasing $M$ we should approach the optimum. For the homogeneous contrast case, the most promising are those having the smallest "number of '1' in light areas + number of '0' in dark areas" - the best possible contrast is as for the lowest achievable grayness.

Finally the problem is: we have an ensemble of $M$ values (the amount of '1's) so far, to all of them we separately add $2^f$ independent variables from ${n \choose i}/2^n$ binomial distribution, then only $M$ minimal of these $2^f M$ values survive and so on. The question is the minimal grayness: how fast the minimum in the ensemble grows this way? The dependence from $M$ seems difficult to find analytically, but can be easily find in simulations - from Fig. \ref{comp} we see that for any $f$ to get 2\% from the optimal contrast we should use $M\approx 30$, or $M\approx 300$ for 1\% difference.

This $M$ is linear coefficient of time and memory complexity of encoding - is paid only once for given code and 2D codes are relatively small, so even for very large $M$ this cost is practically negligible. The necessity of choosing the smallest $M$ possibilities suggests the complexity grows rather like $M \lg(M)$ to sort them, but as small variance of $M$ is not a problem, we could just use some fixed number of buckets for weight ranges.
\section{Kuznetsov and Tsybakov case}
The Correction Trees approach to error correction works perfectly if errors are distributed uniformly - in other cases, rarely there can appear very costly to correct error concentrations - the worse, the less frequent. Situation in currently discussed purpose is much better: there is not only a single satisfying path (the proper correction), but there should be statistically essential population of them. However, rare local concentrations of constrains still require essentially larger number of steps - we will now try to understand these issues from constrained coding point of view.
\begin{figure}[b!]
    \centering
        \includegraphics{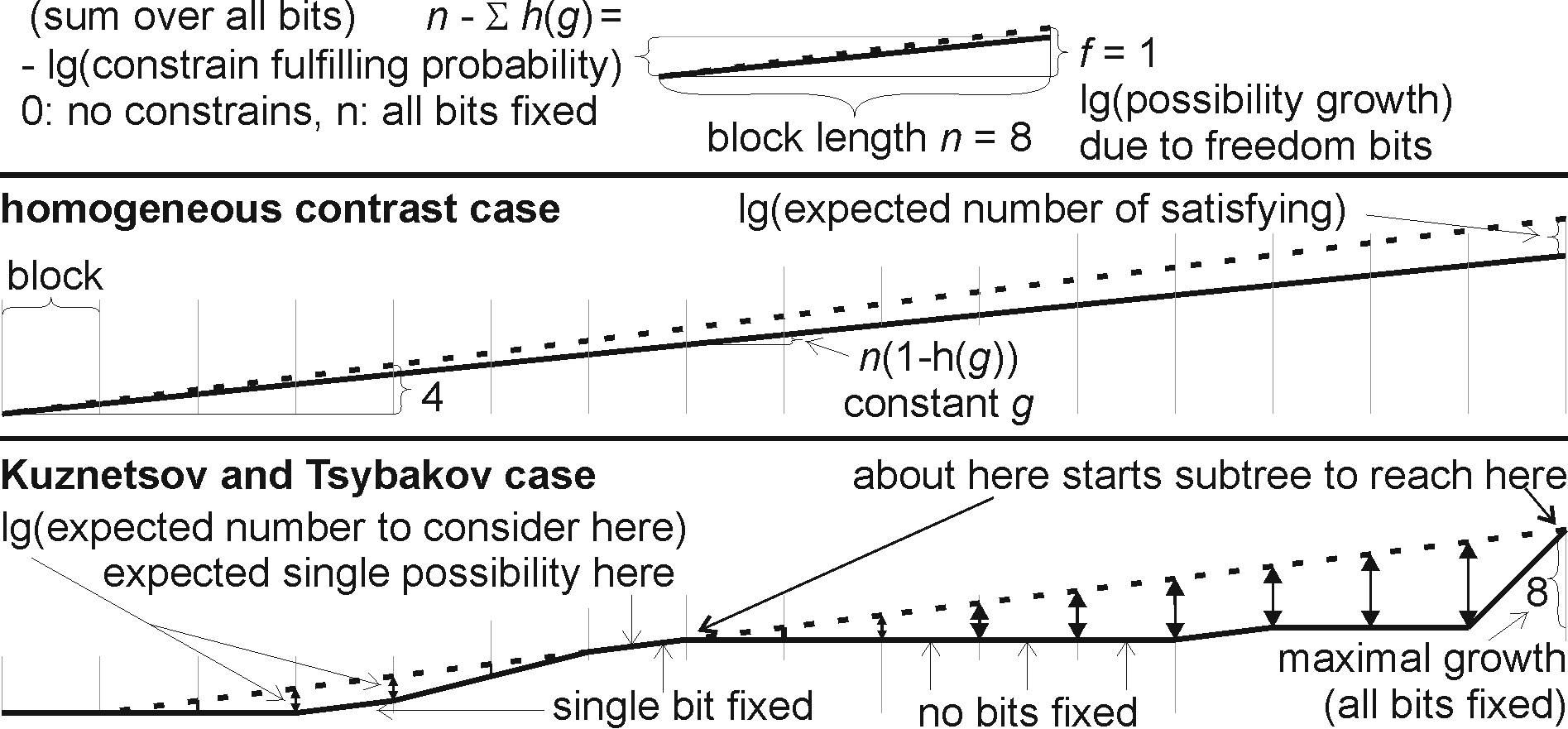}
        \caption{Top: the dashed line corresponds to constant growth of freedom of possibilities and should be above the solid line representing probability of fulfilling the constrains. Center: minus logarithm of probability of fulfilling the constrains grows linearly in homogeneous case. Bottom: it grows per block by the number of fixed bits in this block for Kuznetsov and Tsybakov case. This kind of plot for specific constrains (grayscale picture) allows to initially estimate the number of required steps to find satisfying code.}
       \label{hom}
\end{figure}

Let us start with the Kuznetsov and Tsybakov case: some bits are fixed and the rest of them can be freely chosen. While searching for constrained code in homogeneous contrast case is kind of similar to correcting black and white GP for Binary Symmetric Channel, the current scenario corresponds to Erasure Channel: searching for the constrained code can be seen as searching for a correction where the fixed bits are known and the rest of them are erased.

This time evolving ensemble approach is ineffective: instead of some order among the most promising possibilities, now possibilities just satisfy the constrains or not. So it is enough to develop a single path at a time, returning to an earlier branch if needed - exactly like in the sequential decoding discussed in \cite{cortre}. The situation looks like in Fig. \ref{hom} - the freedom allows us to increase the number of possibilities $2^f$ times per block, while every fixed bit reduces twice the expected number of survived possibilities. So the expected number of possibilities considered in given position is the (base 2) exponent of the difference between drawn lines - some rare local constrain concentrations may be very costly. It is the lesson to wisely choose the pixel ordering to reduce probability of such concentrations. Creating such plot allows to quickly estimate the expected total number of required steps for given constrains by summing these exponents - allowing to try to reduce this number before the search, e.g. by a more convenient pattern positioning or using different ordering from a few available in given standard.

\begin{figure}[b!]
    \centering
        \includegraphics{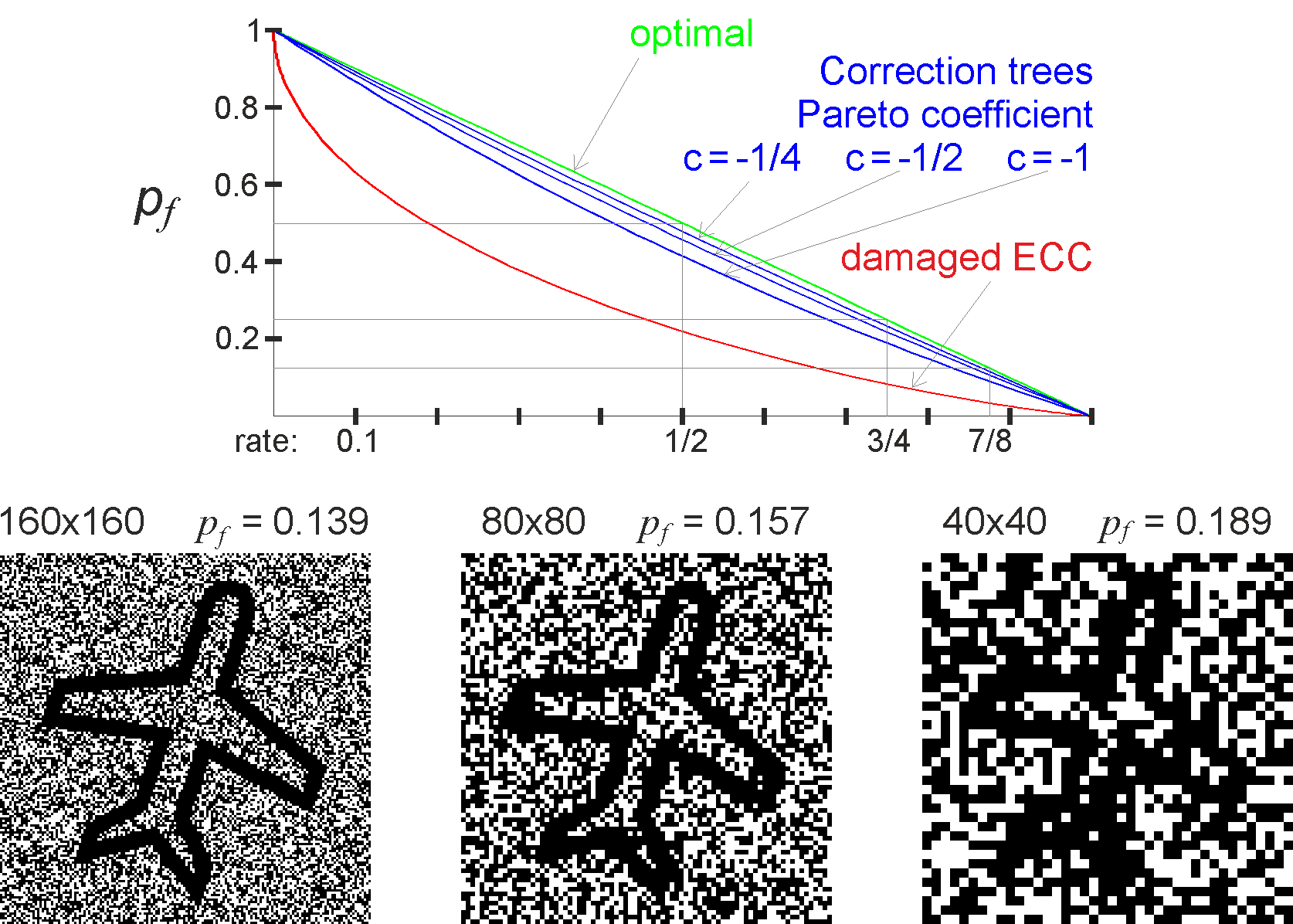}
        \caption{Limits and examples for the Kuznetsov and Tsybakov case. Pareto coefficient $c$ means that probability of requiring to consider more than $m$ possibilities in a single step is approximately $m^c$ (times a constant which occurs to be nearly 1).}
       \label{kt}
\end{figure}

Even if we properly choose pixel ordering to make constrains nearly uniformly distributed in obtained bit sequence, still rarely there can accidently appear very costly concentrations. Using sequential decoding this time makes we get analogous (kind of dual) considerations as for the Erasure Code from appendix of  \cite{cortre}: defining $p_f$ as independent probability that given bit will be fixed, it corresponds to $1-p_e$ of Erasure Codes. Looking at Fig. \ref{hom}, we see this time the current expected number of possibilities (exponent of difference between solid and dashed line) depends on the further situation - to find probability of requiring large number of steps, it is more convenient to make steps backward in contrast to Erasure Codes considerations. As in that analysis, let us start with imagining 1 bit blocks, still containing $q=f/n\in[0,1]$ freedom bits - for analysis it means that the number of possibilities grows $2^q$ times per block. Such fractional amount of bits is technically difficult to realize here, but will allow to understand the general case.

So let us define $T(s)$ as the probability that the expected number of possibilities is smaller than $2^s$ and express it using situation in the succeeding position:
\be T(s)=\left\{\begin{array}{ll} p_f T(s - q + 1) + (1-p_f) T(s-q) \qquad & \textrm{for }s\geq 0\\
                 0 & \textrm{for }s<0 \end{array}\right. \label{vequ}\ee
Assuming $c$ coefficient Pareto distribution as asymptotic behavior and substituting:
\be 1-T(s)\propto 2 ^ {c s} \label{par} \ee
$$2 ^ {c s} =p_f 2^{c(s-q+1)}+(1-p_f)2^{c(s-q)}$$
\be 2^{c q}=2^{c}p_f+1-p_f\qquad\qquad 2^{c f}=(2^{c}p_f+(1-p_f))^n \label{veq}\ee
In the general block size case, the functional equation (\ref{vequ}) contains $2^n$ terms, which leads exactly to power expansion of the right formula (\ref{veq}). We can use this final implicit formula to find the Pareto exponent $c$ for given $p_f$ and $q=f/n$, like in Fig. \ref{kt}. It allows to estimate the probability that the expected number of steps will exceed time and memory resources we would like to use. Implementation of Correction Trees considers a few millions of possibilities per second on modern personal computer, so in practical applications we can use $c$ between $-1/2$ and $-1/4$, which for example for rate 1/2 corresponds to $p_f$ between about 0.46 and 0.48, what is close to the theoretical limit 0.5 corresponding to $c=0$.
\section{The general case (AWGN analogue)}
Combining two previous sections we can use 5 different graynesses: some bits are fixed (grayness 0 or 1), some without constrains (grayness 1/2) and finally the rest have maximal possible contrast toward lightness or darkness. The last group kind of takes the remaining freedom - the larger it is, the better the contrast. This combination rather requires the evolving ensemble approach, but for any fixed size of ensemble ($M$), the randomly distributed fixed bits can rarely make that the whole ensemble will die out ((\ref{par}) Pareto distribution). To prevent going back in such cases, it should be enough to vary $M$ accordingly to the expected number of possibilities to consider - using plot constructed like in Fig. \ref{hom}: choose $M$ as some $M_0$ times base 2 exponent of the difference between solid and dashed lines from such plot. The analysis of situation becomes even more complicated than for the homogeneous contrast case, but the fact that the right plot from Fig. \ref{comp} is nearly independent of $f$, suggests to use this plot to choose satisfying $M_0$, what is confirmed by simulations.\\

Let us get to the main question: how to generalize this approach to allow for any varying grayness to get codes like in Fig. \ref{lena}? We would initially want some pixels to be rather black (if $g>1/2$) or white (if $g<1/2$), but the grayness determines the probability of allowing this color to be changed:
\be [0,1/2]\ni\ \ \epsilon:=1/2-|1/2-g| \qquad\qquad (=g \textrm{ for } g\leq 1/2,\ \ 1-g \textrm{ for }g>1/2)\ee
making it similar to the Additive White Gaussian Noise (AWGN) correction: grayness corresponds to soft information. The survival in the ensemble was previously determined by having the least bits changed from given black and white picture - now we need to use some weights to make that changes of some bits are more acceptable. Changing a fixed bit ($g=\pm 1$) is completely unacceptable: can be imagined that this possibility gets $+\infty$ weight, what automatically takes it out of the ensemble. In the correction problem we use bayessian analysis to find the most probable possibility(leaf) accordingly to already created tree - for constant length path we should just choose the most probable paths assuming expected probability distribution (grayscale picture). To translate it to weight minimization to generalize the previous evolving ensemble approach, we can take minus logarithm of this probability (Fano metric \cite{fano}). Finally this weight for minimization is the sum over all processed bits of
\be \Delta w = \left\{\begin{array}{ll} -\lg(1-\epsilon) & \textrm{if color is as expected: white for $g<1/2$, black for $g\geq 1/2$}\\
                 -\lg(\epsilon) & \textrm{if color is not as expected}\end{array}\right. \label{awgn}\ee
The second possibility should statistically happen in $\epsilon$ cases, so the average growth of weight in this position is $-\epsilon \lg(\epsilon)-(1-\epsilon)\lg(1-\epsilon)=h(\epsilon)=h(g)$ as expected - to initially estimate the behavior by creating plot like in Fig. \ref{hom}, the solid line should grow by $h(g)$ per bit of grayness $g$. While building the tree, the weight is the sum of father's weight and weight of the current bit block.

Finally the algorithm is considering ensemble (e.g. of varying size like in the beginning of this section): expand all paths a single step in all possible ways, updating their weights accordingly to (\ref{awgn}), then choose some number of those having the smallest weight in this new population and so on.

Some generated codes might be not visually satisfactory - we can use some other from the final ensemble instead. This algorithm requires that the rate is indeed approximately the average of $h(g)$ over all pixels. Violating this condition would shift the grayness toward 0.5 or 0/1, decreasing or increasing contrast in nonlinear way. Fulfilling this condition might require not natural $f$ - it can be achieved by varying $f$ between blocks. Simpler way is just to initially modify the picture in controlled way to make we require natural $f$.\\

\begin{figure}[t!]
    \centering
        \includegraphics{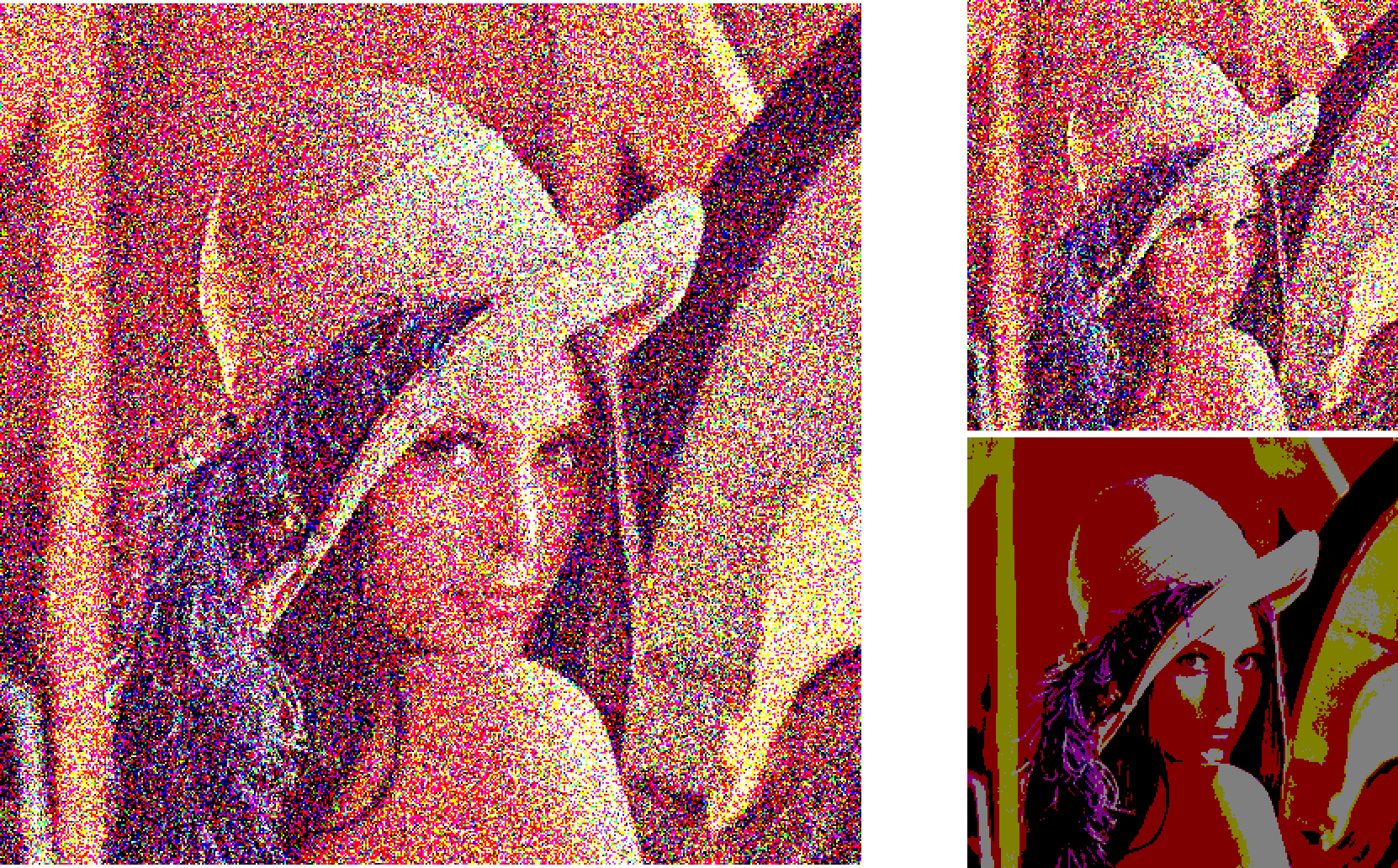}
        \caption{Examples of 3bit/pixel color analogs of Fig. \ref{lena} - this time levels of three basic colors of RGB picture are treated independently as they would be grayscale levels. The visual aspect also reduces the rate to about 0.83, so the left code may contain about 79kB of 96kB required to write it - storing halftone picture alone costs about 18kB (Section \ref{ratdis}).}
       \label{clena}
\end{figure}

This application can be seen as steganography with extremely small number of bits per pixel, like 1 for black and white or 3 for color pictures (Fig. \ref{lena} and \ref{clena}). In such cases, we cannot just distort the least important bits as usually, but encode in the freedom of grayness realization instead. We can also think of intermediate scenarios: if the staganographic picture has to contain less amount of bits per pixel than the original picture. In such a case we can use the original bit values for all but the least important one, for which this time the grayness is determined by less important bits of the original picture - we can now reproduce this grayness, still encoding information in this bit.

\section{Combining with error correction}
The initial purpose of Correction Trees was error correction as enhanced concept of Convolutional Codes, among others by using much larger states (64 bits instead of about 8), what requires replacing the convolution with carefully designed coding procedure - finally making it alternative to modern state of arts methods. It has also essential advantages, like that up to some medium noise level the correction is nearly cost free. Another advantage is that in comparison to Turbo Codes and Low Density Parity Check, this time we have much better control of the correction process, allowing for much larger flexibility of considered damage space, what is extremely important while complex correction of information contained in pictures.

Before discussing the use of this flexibility, let us adapt/modify the original coding from \cite{cortre} to split error correction (currently used also to modify the code in extremely inefficient - damaged ECC way), into two optimized separate purposes: error correction and constrained coding. It can be easily obtained by constructing the preprocessed block ($x$) from all three types of bits: $k$ payload bits, $f$ freedom bits and $R=n-k-f$ redundancy bits (the final rate is $k/n$). Now the freedom bits can be freely chosen while encoding and are discarded while decoding. The redundancy bits are e.g. fixed to '0' while encoding and are used only while correction - obtaining '1' there denotes that we are on a wrong path. Probability that a wrong correction will survive a single step by accidently generating the proper redundancy bits (regularly distributed checksum) is $p_d=2^{-R}$. Finally, the encoder builds one tree using freedom to fulfill the constrains, while the decoder builds a different tree using the redundancy if correction is required.

Another required modification of the original coding is that this time we rather require nonsystematic one: for this purpose the transition table from the original coding should transform the whole symbol ($t: [0,2^n-1] \to [0,2^N-1]$, where $N$ is the size of the state) and the cyclic rotation of state is by at least $n$ (should not divide the size of the state). Now the redundancy check is that XOR of $y$ and the corresponding bits of $s$ is in the proper subset of $[0,2^n-1]$. If not, we can easily determine the nearest corrections (e.g. tabled). For error correction purposes, this subset should be chosen to maximize Hamming distance between allowed sequences, like using only even number of '1' in $R=1$ case.

This nonsystematic case with blocks connected by a hidden state is also perfect for cryptographic applications to hide encrypted information in halftone grayness of a picture. For this purposes $t$ can be chosen using a pseudorandom number generator initialized with a cryptographic key.\\

Let us now briefly discuss the correction process in case of 2D codes. We can simultaneously try to develop different possible ways to decode by including them in the list of possibilities to consider (with initial weight as minus logarithm of probability). Now while searching the space of corrections, the weight of only one of these possibilities will statistically grow (the proper one), while the rest of them will be quickly dominated. It makes it is unnecessary to directly store basic information like block size or freedom/redundancy level - we can try correcting for all such possibilities and only single one will survive. Characteristic redundant squares in current QR codes to determine direction and orientation are also unnecessary while using Correction Trees - again we can start with all 8 possibilities and only the proper one will survive.

Another issue is the pixel ordering which should be chosen to statistically distribute constrains/damages in nearly uniform way in the bit sequence. It can be made by using 2D shifts modulo the dimensions of the code, with carefully chosen shifts - to make plots like in Fig. \ref{hom} relatively flat for more probable shapes. If this plot accidently occurs to be inconvenient for the specific constrains, we could try modifications e.g. by shifting the pattern. Another possibility is using a few different bit orderings - the encoder chooses the most convenient one, while decoder try out all of them and only one survives.

Going to the proper correction, besides considering the standard BSC damages, the correcting algorithm can for example online observe that errors are concentrated in some location and use this information to localize further damages. It can also consider much more complex errors, like shifted or glued pixels, or even try out a space of possible local deformations caused e.g. by folding the medium. We could also try to get out of restriction to black and white codes - the problem with different illumination/perception of colors can be again handled by trying out many different possibilities, from which only one should survive.

To improve the initial search for the proper parameters and damage map, there can be used more redundancy ($R$) in the beginning of our sequence. However, there is some additional information this approach rather requires: the final state. It is absolutely necessary for improving performance by bidirectional correction, while for unidirectional it can be omitted at cost of possibility of not repairing some final bits. This state can be extracted after determining basic information and most of damages - can be stored e.g. as the last bits in our pixel ordering.
\section{Rate distortion application}
\label{ratdis}
Imagine fixing the message to send as just zeroes (or some arbitrarily fixed values) and we do not discard the found freedom bits, but store them instead. These bits are now enough to decode the picture, but in distorted/halftone version. For example encoding this way the halftone pictures alone from Fig. \ref{ex} would require only correspondingly 1/8, 1/4 or 1/2 bits required to encode the black and white picture.

This kind of lossy compression is the standard rate distortion scenario - in which for given rate we search for the closest achievable encoding of some message. We need to define the metric of what being close means - for homogeneous contrast case it is the average number of flipped bits (Hamming distance).

Kuznetsov and Tsybakov case does not fit into this nomenclature: it would correspond to metric growing from zero to infinity if only some of the fixed bits were changed. However, this exotic version of rate distortion could be also practical: if in $n$ bit message we would like to fix only some $k<n$ bits (receiver don't know which, the rest of bits are random), it is enough to send a bit more than $k$ bits.

The general case considered here (AWGN analogue) can be seen as rate distortion using metric with weights depending on position. These weights define how important is that given bits remains unchanged. Using Correction Trees they can be even defined online accordingly to context while searching for close coding, for example using some psychoacoustic model estimating importance for human receiver of given bits. Another application could be just encoding "visual aspects" like of Fig. \ref{lena} - instead of using a few bits/pixel to encode given grayscale picture, use only about 0.18 bits/pixel to encode its halftone version (or e.g. 3 times more for color picture).\\

To realize such rate distortion-like applications, we could indeed fix payload bits e.g. to 0, search for satisfying freedom bits like in Sections 2-5, and then store or send this found bit sequence. The considerations and calculations of these sections remain the same, with the only difference of exchanging bits to discard with the essential ones - the rate is "1 minus the original rate". In practice, instead of using zeroes for payload bits, we can just simplify the encoder, like using transition function $t:[0,2^f-1]\rightarrow [0,2^N-1]$. These original payload bits can be also some fixed bit sequence, like a secret shared between both sides for cryptographic purposes. We could also fix only some of these bits, combining rate distortion with constrained coding.
\section{Conclusions}
There was discussed constrained coding as generalization of Kuznetsov and Tsybakov problem of optimally using a channel with constrains known only to the sender, to include also statistical constrains, like making the code resemble chosen grayscale picture. While fixing some bits could be applied only to a part of the code, these weaker constrains can be applied to all bits simultaneously. Another applications could be optimally using a channel with varying preferences or sending information in faking some varying noise - especially if only the sender can effectively estimate these variations.

The direct application from which perspective this paper was written - 2D codes, besides providing natural intuitions for presented considerations, can lead to the next generation of such codes - optimized to contain also direct visual information for human receiver. For example a code looking like a singer could contain a music sample, some informatively looking code could contain additional information about an item in store or museum, code looking like a logo of operating system could contain a small application (or a virus) related to the specific location.

From Fig. \ref{lena} we see that codes looking like grayness pictures require rather large resolution, like 150x150 for a face - it would allow for larger capacity and so more interesting content, but is also much more demanding from error correction point of view. As discussed in the previous section, Correction Trees also allows for large flexibility of the correction process, what allows for larger capacity codes and to omit visible redundancy. We could also construct multi-level codes - for example poor quality photography would allow to decode some most essential part of the information stored e.g. in grayness of larger blocks, while good quality photography would provide the whole contents.

This approach can be also directly used for steganographic purposes: hiding information in freedom of choosing grayness when standard techniques are unavailable, like for black and white pictures or when large capacity is required - even using only 1 bit/pixel, it still can contain surprisingly large amount of information, like 512x512 pixel picture directly contains 32kB, while making it "looking like \emph{Lena}" costs only about 6kB.

The previous section shows that presented considerations are kind of dual to rate distortion for lossy compression. Besides these additional applications, we could also use it to extremely cheaply encode halftone pictures or long messages with only some bits fixed.

Another way to average pixels to get impression of grayscale from black and white pixels is time average. For example creating animation from many codes looking like the same picture, we would get better impression of the grayscale picture, by the way transmitting a large amount of information.

\bibliographystyle{plain}

\begin{thebibliography}{4}
\bibitem{init} A. V. Kuznetsov and B. S. Tsybakov, \emph{Coding in a memory with defective cells}, Probl. Peredachi Inf., 10:52–60 (1974),
\bibitem{qr} ISO 18004:2000, Information technology -- automatic identification and data capture techniques -- bar code symbology -- QR code, ISO, Geneva, Switzerland,
\bibitem{me} J. Duda, \emph{Asymmetric Numeral Systems}, arXiv: 0902.0271 (2009),
\bibitem{cortre} J. Duda, P. Korus, \emph{Correction Trees as an Alternative to Turbo Codes and Low Density Parity Check Codes}, arXiv: 1204.5317 (2012),
\bibitem{cortre1} J. Duda, http://demonstrations.wolfram.com/CorrectionTrees/,
\bibitem{steg} J. Fridrich, M. Goljan and D. Soukal, \textit{Steganography via codes for memory with defective cells}, Proceedings of the Forty-Third Annual Allerton Conference On Communication, Control and Computing (2005),
\bibitem{fano} R. M. Fano, \emph{A heuristic discussion of probabilistic decoding}, IEEE Transaction on Information Theory, 9:64–73 (1963),
\end{thebibliography}

\end{document}